\documentstyle[a4wide,12pt]{article}
\begin{document}
\begin{center}
{\Large\bf Spacetime Structure and Quantum Physics}\\
\vspace{1cm}
{\large Vu B. Ho}\\
\medskip {\it Department of Physics}\\{\it Monash University}\\
{\it Clayton, Victoria 3168, Australia}
\end{center}
\vspace{1cm}
\centerline {\bf Abstract}
A geometrical description is given to electromagnetism as a four-dimensional
spacetime structure. Then it is shown that the classical dynamics of a
charged particle may be determined only by the four-vector potential, while
the existence of an electromagnetic field may depend on the topological
structure of the background spacetime so that radiation can be considered as
a topological effect. In particular when the spacetime structure of
electromagnetism is complex, it is possible to find a connection between
spacetime structure and quantum physics by invoking the method of path
integration.

\begin{flushleft} {PACS number: 03.65.Bz} \end{flushleft}
\newpage
\noindent
In the appendix to this paper we show a close relationship
between differential geometry and quantum dynamics of a particle.
Bohr's hypothesis of quantisation can be described in terms of
elementary geometry and topology. The quantum condition posseses a
topological structure in the sense that the quantum number $n$ in the
Bohr-Sommerfeld's quantum condition $\oint pds=nh$ can be considered as
a winding number, which is a topological invariant.
As a consequence of such description of Bohr-Sommerfeld's
quantum condition we may conjecture radiation as a topological effect
which reflects the changing of a topological structure of the
physical system of an atom. A stationary state can then be considered as
a class of paths of the fundamental homotopy group. Since a path
from one class can not be deformed smoothly to another path which is in a
different class, so the process must require a change of the topological
structure of the system in some way. In this paper we will show that the
process of creation of a photon may only occur at those points of spacetime
manifold where the topological structure is not smooth so that the
integrability condition is not satisfied. Even in classical sense,
the dynamics of a charged particle may only be determined by the four-vector
potential and, furthermore, the electromagnetic field strength and the
four-vector potential may not be defined together on a smooth spacetime
structure of the electromagnetic manifold. For this purpose we will need
first to construct a non-Riemannian spacetime structure to describe
electromagnetism.

Consider a mathematical structure of spacetime such that the change
$\delta A^\mu$ in the components of the vector $A^\mu$ under
an infinitesimal parallel displacement of the form [1,2]
\begin{equation}
\delta A^\mu = -\beta \Lambda_\nu A^\mu dx^\nu
\end{equation}
will make the covariant derivatives
\begin{equation}
\nabla_\nu A^\mu =\frac{\partial A^\mu}{\partial x^\nu} +
\beta \Lambda_\nu A^\mu,
\end{equation}
transform like tensors under general coordinate transformations. Here the
quantities $\Lambda_\nu$ are the connections of spacetime manifold which
will be identified with the electromagnetic four-potential, and the
quantity $\beta$ is an appropriate dimensional constant.
It should be mentioned here that we have assumed the electromagnetic
structure be a possible independent structure of the curved spacetime,
which is not considered as an additional structure to some existing
Riemannian spacetime manifold of a gravitational field [3,4]. From the
definition of covariant derivatives we find

\begin{equation}
\nabla_\sigma A^{\mu_1...\mu_m}_{\nu_1...\nu_n} = \frac{\partial
A^{\mu_1...\mu_m}_{\nu_1...\nu_n}}{\partial x^\sigma} + (m-n)
\beta A^{\mu_1...\mu_m}_{\nu_1...\nu_n}.
\end{equation}

Under a general coordinate transformation the connections $\Lambda_\nu$
will transform as follows

\begin{equation}
\beta{\Lambda'}_\mu = \frac{\partial x^\nu}{\partial {x'}^\mu} (\beta
\Lambda_\nu) + \frac{\partial^2 x^\nu}{{\partial {x'}^\mu}{\partial
{x'}^\sigma}} \frac{\partial {x'}^\sigma}{\partial x^\nu}.
\end{equation}
The connections $\Lambda_\mu$ in general are arbitrary functions of the
coordinate variables and they do not form a tensor unless we consider only
linear transformations. If in a particular coordinate system the
connections are identified with the electromagnetic potentials,
then, since the potentials should transform like a vector, the effect
caused by the extra term in the transformed connections is not an
electromagnetic effect but related purely to coordinate transformations.
This is an inertial effect.

In spacetime manifold of electromagnetism, the geometrical object which
plays the role of the Riemannian curvature tensor of curved spacetime of
gravitation will take the familiar form of the electromagnetic field tensor.
Denoting this geometrical object also by $F_{\mu\nu}$, we find

\begin{equation}
F_{\mu\nu} = \frac{\partial \Lambda_\nu}{\partial x^\mu}- \frac{\partial
\Lambda_\mu}{\partial x^\nu}.
\end{equation}
In this form the quantities $F_{\mu\nu}$ automatically satisfy the
following homogeneous relationships

\begin{equation}
\frac{\partial F_{\mu\nu}}{\partial x^\alpha} + \frac{\partial
F_{\nu\alpha}}{\partial x^\mu} + \frac{\partial F_{\alpha\mu}}{\partial
x^\nu} = 0.
\end{equation}

When a metric $g_{\mu\nu}$ can be introduced onto the electromagnetic
manifold of spacetime as in the defining relation
$ds^2 = g_{\mu\nu}dx^\mu dx^\nu$, from the usual requirement [5,6]

\begin{equation}
\nabla_\sigma A_\mu dx^\sigma=g_{\mu\nu}\nabla_\sigma A^\nu dx^\sigma,
\end{equation}
we obtain the following relationship between the metric tensor and the
connections

\begin{equation}
\Lambda_\mu = \frac{g^{\lambda\sigma}}{2\beta} \frac{\partial
g_{\lambda\sigma}}{\partial x^\mu} =
\frac{1}{2\beta g}\frac{\partial g}{\partial x^\mu} =
\frac{1}{\beta} \frac{\partial \ln\sqrt{-g}}{\partial x^\mu},
\end{equation}
where $g=\det(g_{\mu\nu})$. It is noted that under a gauge transformation
\begin{equation}
{\Lambda'}_\mu = \Lambda_\mu + \frac{\partial \chi}{\partial x^\mu},
\end{equation}
by putting $\chi=\ln\sqrt{\sigma}/\beta$, we obtain
\begin{equation}
{\Lambda'}_\mu = \frac{1}{\beta} \frac{\partial
\ln\sqrt{-\sigma g}}{\partial x^\mu}=\frac{1}{2\beta} \frac{1}{(\sigma g)}
\frac{\partial {(\sigma g)}}{\partial x^\mu},
\end{equation}
where here $\chi$, and hence $\sigma$, is arbitrary function of the
coordinate variables. It is seen from (5), (8) and (10) that the
electromagnetic field strength exists only at points where the condition of
integrability is not satisfied. At those points where the determinant of
the metric tensor is a smooth functions or at least twice differentiable
there will be only four-potential without electrogmanetic field strength.
It is the analytical behaviour of the metric tensor that
determines the electromagnetic field but not the value of its determinant.

By following the same procedure as in general relativity to obtain the
gravitational field equations, we can postulate the Maxwell's system
of nonhomogeneous equations to obtain a complete system of field equations
for electromagnetism

\begin{equation}
F^{\mu\nu}{}_{;\nu}=-\kappa J^\mu,
\end{equation}
where $J^\mu$ is the current four-vector and $\kappa$ is an appropriate
dimensional constant. But by the above discussions of the
dependence of the electromagnetic field strength on the topological
structure of the background spacetime, we see that the differential
relationship in this form is not well defined since the electromagnetic
field strength exists only at those points where the condition of
integrability is not satisfied or the spacetime structure is not smooth.
As a consequence the Maxwell's system of field equations of electromagnetism
must be considered as statistical and as distribution equations between
macroscopically defined physical quantities.

The equations of motion of a charged particle in an
electromagnetic spacetime manifold can also be derived from the
requirement that the path of a particle be a geodesic

\begin{equation}
\frac{d^2 x^\mu}{ds^2} + \beta \Lambda_\nu \frac{dx^\nu}{ds}
\frac{dx^\mu}{ds} = 0.
\end{equation}
These equations of motion can be rewritten as follows

\begin{equation}
\frac{dx^\mu}{ds}=x_0^\mu\exp{\left (\beta\int\Lambda_\nu dx^\nu\right )},
\end{equation}
where ${x_0}^\mu$ are the velocity components of the particle in the
case when there is no field present, i.e. when the spacetime structure is
Minkowskian. In particular when we can obtain the following relationship

\begin{equation}
\beta \Lambda_\mu \frac{dx^\nu}{ds} = -\frac{q}{m} F^\nu_\mu,
\end{equation}
then we regain the Lorentz force law of a charged particle in an
electromagnetic field of classical electrodynamics

\begin{equation}
\frac{d^2x^\mu}{ds^2} = \frac{q}{m}F^\nu_\mu \frac{dx^\nu}{ds}.
\end{equation}

Now we want to show that from the equations of motion of a
charged particle in an electromagnetic curved spacetime we are able to
derive the familiar Coulomb's law of force of electrostatics. By putting
\begin{equation}
2\beta\Lambda_\mu \frac{dx^\mu}{ds} = \eta,
\end{equation}
from (10) we find
\begin{equation}
g=g_0\exp {\left ( \int \eta ds \right )},
\end{equation}
where $\eta$ is some arbitrary function of the coordinates. In the case
when (16) is a homogeneous linear first integral of the equations
of motion then $\eta$ is a constant. In the non-relativistic limit the
determinant (17) can be put in the form
\begin{equation}
g=g_0\exp{\left ( 2\alpha\frac{ct}{r} \right )},
\end{equation}
where $g_0$ and $\alpha$ are constants. With this form of the determinant
of the metric tensor, we then obtain, by defining
$\Lambda_\mu=(\phi,-{\bf A})$, the following four-vector potential

\begin{equation}
\phi = \frac{\alpha}{\beta}\frac{1}{r}, \ \ \ \
{\bf A} = \frac{\alpha ct}{\beta} \frac{{\bf r}}{r^3}.
\end{equation}

{}From the above form of the four-potential we note that at finite time $t$,
except for the origin $r=0$, the electromagnetic field strength defined
in terms of the four-potential by the relationship of (5) vanishes
everywhere. Also with this form of the four-potential we are led to suggest
that there be a charged particle located at the origin where the
electromagnetic field may be different from zero. Any possible influence of
such charged particle on other charged particles can only be via the
potentials themselves.

In the non-relativistic limit the equations of motion (14) take the form
\begin{equation}
\frac{d^2 {\bf r}}{dt^2} + \beta c\phi \frac{d{\bf r}}{dt} = 0.
\end{equation}
Using the potential $\phi$ in (19) we then have
\begin{equation}
\frac{d^2 {\bf r}}{dt^2} + \frac{c\alpha}{r}\frac{d{\bf r}}{dt}=0.
\end{equation}
In the appendix we show that the following semi-classical relation can
be derived from differential geometry and de Broglie's relationship between
the momentum and the wavelength of a particle,
\begin{equation}
\frac{d{\bf r}}{dt}=\frac{\hbar}{m}\frac{{\bf r}}{r^2}.
\end{equation}
The equations of motion then take the familiar Newton's equation of motion
for a test charged particle in spherically symmetric electrostatic field
\begin{equation}
m\frac{d^2 {\bf r}}{dt^2} = -e^2 \frac{{\bf r}}{r^3},
\end{equation}
where we have put $\alpha=e^2/\hbar c$, with $e$ is the fundamental
charge. The constant $\alpha$ is precisely the fine structure
constant, and the constant $\beta$ is equal to $1/\hbar c$. It should be
emphasized here that the definition of the electric field strength in this
case can only be regarded as experimental while the physical quantities
introduced just for the purpose of classical dynamical description.

We have remarked that the determinant of a metric tensor on an
electromagnetic spacetime manifold has the property that it generates
electromagnetic field by its analytical behaviours but not by its values.
This property is reflected through the gauge transformation (10). The result
may lead us to
speculate that there may be some intrinsic relationship between the
determinant of the metric tensor and the wavefunction in quantum mechanics.
Each quantity $g$ corresponds to an infinite number of possible
spacetime structures and the same electromagnetic field can be generated by
all determinants of the form $\sigma g$ which all behave analytically
equivalently. With these remarks we now consider complex structure of
an electromagnetic spacetime where the change of a vector under an
infinitesimal displacement is purely imaginary

\begin{equation}
\delta A^\mu = -i\beta \Lambda_\nu A^\mu dx^\nu.
\end{equation}
The covariant derivatives will take the form
\begin{equation}
\nabla_\nu A^\mu =\frac{\partial A^\mu}{\partial x^\nu} +
i\beta \Lambda_\nu A^\mu.
\end{equation}
We see that with these covariant derivatives we regain the minimal coupling
of a charged particle in an electromagnetic field as gauge description in
quantum mechanics [7,8,9]. But there is a fundamental difference between
the present description and that in quantum mechanics since here we can
consider the Schrodinger equation for a charged particle
in an electromagnetic field as an equation for a free particle in a curved
electromagnetic spacetime.

The determinant $g$ of a metric tensor in an electromagnetic complex
structure can be put in the form

\begin{equation}
g = \exp { \left ( i\int \eta ds \right ) },
\end{equation}
where we have used the gauge condition to set $g_0=1$.
With this form of the determinant of the metric tensor,
in the non-relativistic limit we can identify it with the phase factor in
Feynman path integral formulation of quantum mechanics [10]. However there
is a fundamental difference between the present problem and that of
Feynman's. In the present case a particle is assumed to move along
only one geodesic between two points for a given spacetime
structure, and we have now an infinite number of possible spacetime
structures, unlike that in Feynman path integral
method where we consider a fixed spacetime structure and study the
dynamics of the particle and assume that the particle can take any
path to move from one place to another. This result
may have some relationship with the idea of many-worlds interpretation of
quantum mechanics [11].

We have described some possible connections between spacetime structure
of electromagnetism and quantum dynamics of a particle. However we have not
discussed any kind of relationship between the present formulation of
electromagnetism and strong and weak interactions in particle physics. The
obvious generalisation would be a proper definition of connections for a
spacetime structure to describe classical versions of electroweak field and
quantum chromodynamics.

\section*{Acknowledgments}

I would like to thank Dr. M.J. Morgan for useful discussions on topological
methods in physics. I also would like to acknowledge the financial support of
APA Research Award.

\section*{Appendix}
\setcounter{equation}{0}
\renewcommand{\theequation}{A.\arabic{equation}}

\noindent
In this appendix we will verify the semi-classical relation (22) using
differential geometry and de Broglie's relationship between the momentum
and the wavelength of a particle. Let the path of a particle be represented
by the position vector
${\bf r}(s)$ with the arclength $s$ as a parameter. If ${\bf t}(s)$ and
${\bf p}(s)$ are the unit tangent vector and the unit principal normal
vector of the path respectively, then the formulae of Frenet in the plane
can be written as follows [12]

\begin{equation}
\frac{d{\bf t}}{ds}=\kappa {\bf p},\ \ \ \
\frac{d{\bf p}}{ds}=-\kappa {\bf t},
\end{equation}
where here $\kappa$ is the curvature of the path. By differentiation we
obtain the following differential equation for ${\bf t}(s)$, and a similar
equation for ${\bf p}(s)$,

\begin{equation}
\frac{d^2{\bf t}}{ds^2}-\frac{d(\ln\kappa)}{ds} \frac{d{\bf t}}{ds} +
\kappa^2 {\bf t} = 0.
\end{equation}

If on the curve ${\bf r}(s)$ the curvature $\kappa(s)$ changes slowly so
that in general we can consider the condition $d(\ln\kappa)/ds=0$ always
satisfied, then ${\bf t}(s)$ and ${\bf p}(s)$ can be considered as being
oscillating with a spatial period or wavelength $\lambda$ whose
relationship with the curvature $\kappa$ is of the following form
\begin{equation}
\kappa=\frac{2\pi}{\lambda}.
\end{equation}

Now if we assume that the particle behaves as a wave as in quantum
mechanics so that the wavelength $\lambda$ can be identified with de
Broglie's wavelength of the particle, then the relationship between
the momentum $p$ and the curvature $\kappa$ of the form
\begin{equation}
p=\hbar\kappa
\end{equation}
can be considered as true. This is the magnitude relationship of the
semi-classical vector equation (22).
We know that directional properties of the motion of a
particle in classical sense are not applied in quantum theory due to the
Heisenberg's uncertainty relation. So the vector relationship
as in the semi-classical relation (22) must be regarded as an assumption
on the average of dynamical effects so that in classical mechanics the
dynamics of the particle can be described deterministically.
Due to spherical symmetry of the problem this assumption seems reasonable.
To make the relationship between the momentum and the curvature (A.4) more
plausible, we will show that the Bohr-Sommerfeld's quantum condition and
Feynman's postulate of arbitrariness of the path of a particle follow
naturally from it. The action integral $I=\int pds$ in this case takes the
form

\begin{equation}
I= \int \hbar \kappa ds = \hbar\int\frac{f''}{1+{f'}^2}dx,
\end{equation}
for any path $f(x)$ in a plane we have $\kappa=f''/(1+{f'}^2)^{3/2}$.
By means of the calculus of variations [13], we know that to extremise
the integral $I=\int L(f,f',f'',x)dx$, the function $f(x)$ must satisfy
the following variational differential equation
\begin{equation}
\frac{\partial L}{\partial f} - \frac{d}{dx} \frac{\partial L}{\partial f'}
+ \frac{d^2}{dx^2} \frac{\partial L}{\partial f''} = 0.
\end{equation}
But with the functional $L$ of the form $L=\hbar f''/(1+{f'}^2)$, it is
straightforward to verify that the above variational differential equation
is satisfied by any function $f(x)$. This result can be considered as an
alternative statement of Feynman's postulate of arbitrariness of the path of
a particle in the path integral formulation of quantum mechanics in a
plane. The result also shows that at the quantum level the principle of
least action may break down entirely. Since with the identification of the
momentum with the curvature of the path of a particle, a particle can take
any path to move from one place to another in the plane, then if we choose
circular paths as representatives in each class of paths in the fundamental
homotopy group, we obtain immediately the Bohr-Sommerfeld's quantum condition

\begin{equation}
\oint pds=\hbar \oint \frac{d\theta}{2\pi}=n\hbar.
\end{equation}

In three dimensional space besides path integral we can also
formulate surface integral. We then arrive at the idea of
sum over random surfaces which can be used to construct the quantum
geometry of strings as an alternative procedure for string quantisation
[14,15]. In the following we will discuss the more general idea of sum over
random hypersurfaces. For this purpose let us generalise the action
integral (A.5) by defining an action integral of the following form

\begin{equation}
I_n = \int q_n K_n dS_n,
\end{equation}
where $q_n$ is a universal constant which may be identified with some
physical quantity, depending on the dimension of space. $K_n$ is the
generalised Gaussian curvature [16] defined as the product of the
principal
curvatures $K_n=k_1k_2...k_n$. In terms of Riemannian curvature
tensor the Gaussian curvature can be written in the following form

\begin{equation}
K_n =\frac{1}{2^{\frac{n}{2}}n!det(g_{\mu\nu})} \epsilon^{\mu_1...\mu_n}
\epsilon^{\nu_1...\nu_2}R_{\mu_1\mu_2\nu_1\nu_2}... R_{\mu_{n-
1}\mu_n\nu_{n-1}\nu_n} ,
\end{equation}
where $\mu,\nu=1,2,...,n$.
The following variational differential equations can be derived from the
action integral by the calculus of variations [13,17]

\begin{equation}
\frac{\partial L}{\partial f} + \frac{\partial}{\partial x^\mu}
\frac{\partial L}{\partial f_\mu} + \frac{\partial^2}{{\partial x^\mu}
{\partial x^\nu}}\frac{\partial L}{\partial f_{\mu\nu}} = 0 ,
\end{equation}
where here f is a hypersurface of the form $x^{n+1}=f(x^1,...,x^n)$, and
$L=L(f,f_\mu,f_{\mu\nu})$ with $f_\mu$ is the
partial derivative of $f$ with respect to $x^\mu$.

In the case of $n=2$, by direct substitution, it can be shown that
the variational differential equation (A.10) can be satisfied by any surface.
By considering the homotopy group of surfaces, the action
integral can be reduced to the familiar Gauss's law in electrodynamics
$\oint qKdS =nq$, where $q=q_2$ now is identified with the charge of a
particle. Hence we may suggest that the charge of a physical system is a
manifestation of topological structure of the system under consideration
and the quantisation of charge is a reflection of this relationship.
Finally it is also noted that in the case of
hypersurfaces of even dimensions the above discussions of the relationship
between the action integral and the variational differential equations that
lead to the idea of sum over random hypersurfaces should be justified
because in this case the action integral is in the form of Euler
characteristic. This result also suggests that it may be possible to discuss
a formalisation of the physics that involves infinite dimensional manifolds.

\newpage
\section*{References}
1. L.P. Eisenhart, {\it Non-Riemannian Geometry} (American Math.
Society, New York, 1968).\\
2. A. Einstein, {\it The Meaning of Relativity} (Menthuen \& Co Ltd,
London, Great Britain, 1956).\\
3. H. Weyl, {\it The Principle of Relativity}
(Dover, New York, 1952).\\
4. W. Pauli, {\it Theory of Relativity} (Dover, New York, 1958).\\
5. L.D. Landau and E.M. Lifshitz, {\it The Classical Theory of Fields}
(Pergamon Press, 1975).\\
6. C.W. Misner, K.S. Thorne and J. . Wheeler, {\it Gravitation} (Freeman,
San Francisco, 1973).\\
7. C.N. Yang and R.L. Mills, Phys. Rev. {\bf 96}, 191(1954).\\
8. R. Utiyama, Phys. Rev. {\bf 101}, 1597(1956).\\
9. I.J.R. Aitchison and A.J.G. Hey, {\it Gauge Theories in Particle
Physics} (Adam Hilger, Bristol, England, 1989).\\
10. R.P. Feynman and A.R. Hibbs, {\it Quantum Mechanics and Path
Integrals} (McGraw-Hill, New York, 1965).\\
11. B.S. DeWitt and N. Graham, {\it The Many-Worlds Interpretation of
Quantum Mechanics} (Princeton University Press, Princeton, New Jersey,
1973).\\
12. E. Kreyszig, {\it Introduction to Differential Geometry and
Riemannian Geometry} (University of Toronto Press, 1968).\\
13. R. Weinstock, {\it Calculus of Variations} (Dover, New York,
1974).\\
14. A.M. Polyakov, Phys. Letts. {\bf 103B}, 207(1981).\\
15. A.M. Polyakov, Phys. Letts. {\bf 103B}, 212(1981).\\
16. M. Spivak, {\it A Comprehensive Introduction to Differential
Geometry} (Publish or Perish, Berkerley, 1979).\\
17. A.O. Barut, {\it Electrodynamics and Classical Theory of Fields
and Particles} (Dover, New York, 1980).

\end{document}